\documentclass[%
 reprint, amsmath,amssymb, aps,showpacs]{revtex4-1}

\usepackage{graphicx}
\usepackage{epsfig}
\begin{document}


\title{Topological phase in a $d_{x^2-y^2}+(p+ip)$ superconductor in presence of 
spin-density-wave}

\author{Amit Gupta}

\author{Debanand Sa}%
\affiliation{%
Department of Physics, Banaras Hindu University, Varanasi-221 005
  \\}%
  
\date{\today}

\begin{abstract}
We consider a mean-field Hamiltonian for a $d_{x^2-y^2}+(p+ip)$ superconductor(SC) in presence of spin-density-wave(SDW) order. This is due to the fact that the non-commutativity 
of any two orders produces the third one. The energy spectrum of such a Hamiltonian is 
shown to be gapped and it yields a topological phase in addition to the conventional one. A phase diagram characterizing different topological phases is construted. The Chern numbers and hence the nature of the topological phases are determined. The edge state spectrum and the possibility of whether the vortex state harbouring the zero modes are discussed.
\end{abstract}

\maketitle



The subject of topological systems in condensed matter is one of the most 
active field of research at present and is developing in rapid 
pace\cite{hasan10,qi11}. Topological phases are characterized by the 
existence of both the gapless edge states as well as the gapped bulk 
states\cite{moore91,nayak96,fradkin98,hatsugai93,thouless82}. 
In order to have topological phase, one ought to have an gap in the energy spectrum  
separating the ground state from the excited states. In such a case, one 
can define a smooth deformation in the Hamiltonian which does not close the bulk gap. 
This is due to the fact that one gapped state can not be deformed into another 
gapped state in a different topological class unless a quantum phase transition 
occurs when the system become gapless. This field has attracted a lot of interest 
due to its wide range of applicability in various areas of condensed matter systems 
such as, quantum Hall effect\cite{thouless82}, 
superconductors\cite{read00,ivanov01,stern04}, 
$Z_2$ topological insulators(spin Hall insulators)\cite{kane05,sheng05,bernevig06} 
etc. The topological phases are crucially 
dependent on some particular symmetries of the system such as time reversal(TR), 
space inversion(SI), particle-hole(PH) and chiral etc. The gapless edge states 
are topologically stable against those perturbations that do not break the symmetries 
of the system. The topologically protected gapless edge states play important role 
in determining the transport properties of the system. The total number of topologically protected edge modes in a given system is associated with the topological numbers 
such as the Thouless-Kohmoto-Nightingale-den Nijs(TKNN) number (the first Chern number) 
for the systems without time reversal symmetry\cite{thouless82,hatsugai93} and the $Z_2$ invariant in case of time reversal invariant systems\cite{kane05}. \\ 

The two-dimensional(2D) topological insulators were theoretically predicted by 
Bernevig et. al.\cite{hughes06,bernevig06} and experimentally observed in 
HgTe/CdTe quantum wells\cite{konig07}. Such an insulator was already proposed by 
Kane and Mele in 2005\cite{kane05}. These quantum states of 
matter belong to a class which is invariant under TR symmetry and the spin-orbit 
(SO) coupling is essential to achieve this. Soon after, it was generalized to 
superconductors and superfluids\cite{roy08,schnyder08,kitaev09,qi09}. In 2D, the classification of topological SC is similar to that of topological insulators. 
For example, the TR breaking SC are classified by an integer 
${\cal N}$\cite{volovik88,read00} similar to that of quantum Hall 
insulators\cite{thouless82} whereas TR invariant SC are represented by a $Z_2$ 
invariant in 2D and 1D\cite{roy08,schnyder08,kitaev09,qi09}. The TR breaking 
topological SC have attracted a lot of attention recently due to their relation to 
non-Abelian statistics\cite{read00,ivanov01} and their potential application to 
topological quantum computation\cite{nayak08}. The nature of the low-energy gapless 
edge states in such systems are non-trivial. They imply fractionlization of 
quasi-particles\cite{lee07} as well. For example, in a vortex core of a spinless 
$p+ip$ SC, the zero mode is described by a Majorana fermion which is 
half of a conventional fermion\cite{kopnin91,read00}. Such a vortex with a 
Majoranan fermion obeys non-Abelian statistics which is crucial for the construction 
of fault-tolerant quantum computers. The existence of zero energy Majorana mode\cite{volovik99} in a vortex core characterizes the topological order in the system.\\

The search for the possible realization of topological phases in condensed matter 
systems is an intriguing and challenging issue. This involves novel concepts as 
well as potential applications. In this communication, we consider a    
coexistence phase of singlet SC and SDW which induces a triplet SC component. 
This is precisely due to the non-commutativity of the former two orders.
The singlet SC is taken to be of $ d_{x^2-y^2}$  symmetry whereas the SDW order 
parameter is of s-wave and the triplet SC is of $p+ip$ type symmetry. Such a Hamiltonian
is shown to yield a non-trivial coexistence phase which is topological in addition to 
the conventional one. A phase diagram characterizing different topological phases is 
construted. The Chern numbers and hence the nature of the topological phases are 
determined. The edge state spectrum and the vortex state in such system are also 
discussed.\\  

\section{Theoretical formulation}
\label{} 
Motivated from the recent spectroscopic experimental results on the appearence of a 
nodal gap on the deeply underdoped cuprate SC\cite{tanaka06,vishik12,ino00,razzoli13,peng13,shen04}, there has been few studies\cite{lu14,das13,gupta15} to uncover such new and unexpected results. In an earlier 
work, we have already discussed about the topological study  
of d-wave SC in presence of SDW order and compared with the above cuprates data\cite{gupta15}.
However, from the study of the group algebra\cite{zhang97,markiewicz98}, it is known that 
the coexistence of any two non-commuting order parameters produces a third order 
parameter. In case of SDW and d-wave superconductivity, there is a third, dynamically generated, order parameter\cite{kyung00}. This happens to be a triplet SC here. In the present work, we consider the coexistence of SDW and d-wave superconductivity which 
can generate a triplet and non-zero center of mass superconducting order parameter. 
We, thus start with a Hamiltonian on a $ 2 $D  square lattice as,  

\begin{widetext}
\begin{eqnarray}
{\cal H}=
\sum_{k, \sigma} \xi_{k}c^{\dagger}_{k,\sigma}c_{k,\sigma}+\frac{U}{N•}\sum_{k,k'}c^{\dagger}_{k,\uparrow}c_{k+Q,\uparrow}c^{\dagger}_{k'-Q,\downarrow}
c_{k',\downarrow} 
+ \sum_{k,k'}V^{1}(k,k')c^{\dagger}_{k,\uparrow}c^{\dagger}_{-k,\downarrow}c_{-k',\downarrow}c_{k',\uparrow}
\nonumber\\
 +\sum_{k,k'}V^{2}(k,k')c^{\dagger}_{k,\uparrow}c^{\dagger}_{-k-Q,\downarrow}
 c_{-k'-Q,\downarrow}c_{k',\uparrow}. 
\end{eqnarray}
\end{widetext}

\noindent Here, $ \xi_{k} $ is the bare dispersion, $U$ is the onsite Coulomb interaction, 
$ V^{1,2} $ are the pairing strengths for $d$-wave and $p$-wave superconductivity 
and $N$ is the number of sites. We model the bare dispersion in the tight-binding approximation on a 2D square lattice as 
$ \xi_{k} =-2t(\cos k_{x} +\cos k_{y}) - 4t'\cos k_{x}\cos k_{y} -\mu $. 
$c^{\dagger}_{{{k}}\sigma}$ ($c_{{{k}}\sigma}$) denotes creation (annihilation) operator 
of the  electron with spin $ \sigma=(\uparrow,\downarrow)$  at ${\textbf{k} }=(k_x,k_y)$. Here, $\sum_{k}^{'}$ is the sum of $k$ over the reduced Brillouin zone (RBZ). 
We express the wave-vector $k$ in units of  $ \frac{\pi}{a} $, with '$a$' the lattice parameter of the underlying square lattice. ${\bf Q} = (\pi, \pi)$ is the SDW nesting 
vector in 2D. We assume here a commensurate 
SDW so that $ {\bf {k + Q}} ={\bf {k-Q}} $. The staggered spin magnetization is defined as $M_0 = -\frac{U}{N}\sum_{k, \sigma}\sigma<c^{\dagger}_{k+Q,\sigma}c_{k,\sigma}>$. Since we will be discussing about three order parameters below, the crystal symmetry of them should 
be such that the commutator of any two of them should give the third one. For this reason, 
if $ V^{1}$ is assumed to be of singlet d-wave symmetry, the SDW state guarantees that 
$ V^{2}$  should be of triplet type. So we get the singlet 
interaction $V^{1}_{k,k'} =V_{0}^{1} s_{k} s_{k'}$ and $V^{2}_{k,k'} =V_{0}^{2} p_{k} p_{k'}$, where $s_{k} =\frac{1}{2}(\cos k_{x} -\cos k_{y}) $ and 
$p_{k} =\sin k_{x} + i\sin k_{y}$. 
We assume that $V^{1,2}$ are attractive. The SC order parameters are defined as, 
for singlet state, $\bigtriangleup^{1}_{k'} = \bigtriangleup^{1}_{0} s_{k'} = V_{0}^{1} s_{k'}\sum_{k}s_{k}<c^{\dagger}_{k,\uparrow}c^{\dagger}_{-k,\downarrow}> =V_{0}^{1} s_{k'}  \sum_{k}^{'}s_{k}<c^{\dagger}_{k+Q,\uparrow}c^{\dagger}_{-k+Q,\downarrow}>$.  On the other hand, the triplet order is decoupled in the main band and in the magnetic band as $\bigtriangleup^{2}_{k'} = \bigtriangleup^{2}_{0} p_{k'} = V_{0}^{2} p_{k'}\sum_{k}p_{k}<c^{\dagger}_{k+Q,\uparrow}c^{\dagger}_{-k,\downarrow}>= \bigtriangleup_{0}^{2}(\sin k_{x} + i\sin k_{y})=\bigtriangleup_{1,k}^{2}+ i\bigtriangleup_{2,k}^{2} $ while $\bigtriangleup^{2\ast}_{k'} = V_{0}^{2} p_{k'}\sum_{k}p_{k}<c^{\dagger}_{k,\uparrow}c^{\dagger}_{-k+Q,\downarrow}>$. This is the reason why the time-reversal 
symmetry remains invariant in the triplet SC state. After substituting these mean-field orders, the total Hamiltonian reads as,

\begin{widetext}
\begin{eqnarray}
{\cal H}=
\sum_{k, \sigma} \xi_{k}c^{\dagger}_{k,\sigma}c_{k,\sigma}+M_{0}\sum_{k,k'}\sigma c^{\dagger}_{k+Q,\sigma}c_{k,\sigma} +\sum_{k}\bigtriangleup^{1}_{k}(c^{\dagger}_{k,\uparrow}c^{\dagger}_{-k,\downarrow}+c_{-k,\downarrow}c_{k,\uparrow})
 +\sum_{k}\bigtriangleup^{2}_{k}c^{\dagger}_{k,\uparrow}c^{\dagger}_{-k-Q,\downarrow}+\bigtriangleup^{2\ast}_{k}c_{-k-Q,\downarrow}c_{k,\uparrow}
\nonumber\\ 
=\sum_{k, \sigma}^{'} [\xi_{k}^{+}(c^{\dagger}_{k,\sigma}c_{k,\sigma}+c^{\dagger}_{k+Q,\sigma}c_{k+Q,\sigma})+\xi_{k}^{-}(c^{\dagger}_{k,\sigma}c_{k,\sigma}-c^{\dagger}_{k+Q,\sigma}c_{k+Q,\sigma})]
+M_{0}\sum_{k,\sigma}^{'}\sigma c^{\dagger}_{k+Q,\sigma}c_{k,\sigma}
\nonumber\\
 +\sum_{k}^{'}[\bigtriangleup^{1}_{k}(c^{\dagger}_{k,\uparrow}c^{\dagger}_{-k,\downarrow}+
 c_{-k,\downarrow}c_{k,\uparrow}-c^{\dagger}_{k+Q,\uparrow}c^{\dagger}_{-k-Q,\downarrow}-
 c_{-k-Q,\downarrow}c_{k+Q,\uparrow})]
\nonumber\\
 +\sum_{k}^{'}[\bigtriangleup^{2}_{k}(c^{\dagger}_{k,\uparrow}c^{\dagger}_{-k-Q,\downarrow}-c^{\dagger}_{k+Q,\uparrow}c^{\dagger}_{-k,\downarrow} )+\bigtriangleup^{2\ast}_{k}
 (c_{-k-Q,\downarrow}c_{k,\uparrow}-c_{-k,\downarrow}c_{k+Q,\uparrow})], 
\end{eqnarray}
\end{widetext}

\noindent where $ \xi_{k}^{+}=- 4t'\cos k_{x}\cos k_{y} -\mu $ and
$ \xi_{k}^{-}=-2t(\cos k_{x} +\cos k_{y}) $.
In the above Hamiltonian, the nesting property in the band dispersion i.e. 
$\xi_{k+Q}^{+}= \xi_{k}^{+} $, $\xi_{k+Q}^{-}= -\xi_{k}^{-}$ and the order parameters  
$\bigtriangleup^{1}_{k+Q} 
=-\Delta_{1}(\frac{\cos k_{x}a-\cos k_ya}{2}) 
=-\bigtriangleup^{1}_{k}$ and 
$ \bigtriangleup^{2}_{k+Q}=-\bigtriangleup^{2}_{k} $ have been employed. 
In the momentum space, the Hamiltonian can be expressed as, 
${\cal H}=\sum_{k}\psi^{\dagger} _{k} {\cal H}({k})\psi_{k}$ 
where the four-component spinor $\psi_{k}$ is,  
$\psi^{\dagger}_{k}=(c_{k\uparrow}^{\dagger},c_{-k-Q\downarrow},
c_{-k\downarrow}^{\dagger},c_{k+Q\uparrow})$.   
Thus, the Hamiltonian matrix  $  {\cal H}({k}) $ in this basis is written as, \\
 
\begin{equation}
\label{eq.2}
{\cal H}({k})= \left( \begin{array}{cccc}
\xi_{k}^{+}+ \xi_{k}^{-}& \bigtriangleup^{2}_{k}  & \bigtriangleup^{1}_{k} &  M_{0}\\
\bigtriangleup^{2\ast}_{k} & -\xi_{k}^{+}+ \xi_{k}^{-} & M_{0}  & -\bigtriangleup^{1}_{k}\\
\bigtriangleup^{1}_{k} &  M_{0} & -(\xi_{k}^{+}+ \xi_{k}^{-})  & -\bigtriangleup^{2\ast}_{k} \\
M_{0} & -\bigtriangleup^{1}_{k}& -\bigtriangleup^{2}_{k} & \xi_{k}^{+}- \xi_{k}^{-}
\end{array} \right).
\end{equation} 

In what follows, we study the energy spectrum of the above Hamiltonian. 
The Hamiltonian (eqn.(3)) is diagonalized and the quasiparticle spectrum is obtained as,

\begin{widetext}
 $ E_{\pm,\pm}(k) 
=\pm\sqrt{\xi_{k}^{+2} +\xi_{k}^{-2}+(\bigtriangleup^{1}_{k})^2 
+ \mid \bigtriangleup^{2}_{k}\mid^{2}+M_{0}^{2}\pm 2 \sqrt{\xi_{k}^{-2}\mid \bigtriangleup^{2}_{k}\mid^{2} +
(\bigtriangleup^{2}_{1,k} \bigtriangleup^{1}_{k}-M_{0}\xi_{k}^{+})^{2}+\xi_{k}^{+2} 
\xi_{k}^{-2} }}$.
\end{widetext}

\begin{figure}
\includegraphics[scale=0.60]{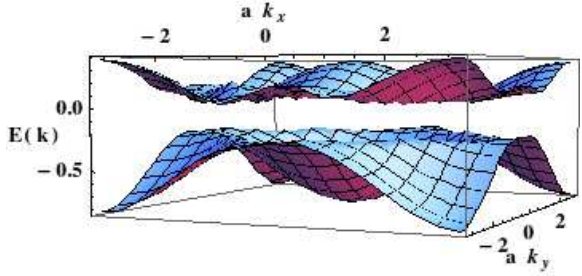} 
\caption{Energy spectra  $ E_{\pm,+}(k)$, corresponding to 
coexistence of SC order parameters $ d_{x^{2}-y^{2}}+(p_{x}+ ip_{y})$ and that of 
the SDW order  parameter showing fully gapped spectrum. For 
illustration, here, we have chosen $ M_0 =0.3t$ eV, $ t'=-0.3t $,
$ \bigtriangleup_{0}^{1}=\bigtriangleup_{0}^{2}= t$, $ \mu=0.25t $ ($t=0.15$ eV.) }
\label{fig.1}
\end{figure}

\noindent It is obvious that the energy spectrum is fully gapped and the gap 
closes only when the following condition is satisfied. 

\begin{equation}
\begin{split}
&\xi_{k}^{+2} +\xi_{k}^{-2}+(\bigtriangleup^{1}_{k})^2 
+ \mid \bigtriangleup^{2}_{k}\mid^{2}+M_{0}^{2}\\
=& 2 \sqrt{\xi_{k}^{-2}\mid \bigtriangleup^{2}_{k}\mid^{2} +
(\bigtriangleup^{2}_{1,k} \bigtriangleup^{1}_{k}-M_{0}\xi_{k}^{+})^{2}+\xi_{k}^{+2} 
\xi_{k}^{-2}}.
\end{split}
\end{equation}

\noindent It is found from the straight forward calculation \cite{sato06} that this 
condition is equivalent to 

\begin{equation}
\begin{split}
\xi_{k}^{+2} +(\bigtriangleup^{1}_{k})^2 
+M_{0}^{2} =\xi_{k}^{-2}+ \mid \bigtriangleup^{2}_{k}\mid^{2}, \\
(\bigtriangleup^{1}_{k}\xi_{k}^{+}+M_{0}\mid \bigtriangleup^{2}_{k}\mid)^{2}
+(\bigtriangleup^{1}_{k})^2 (\mid \bigtriangleup^{2}_{k}\mid-\bigtriangleup^{2}_{2,k})^{2}\\
+2\bigtriangleup^{1}_{k}(\bigtriangleup^{2}_{2,k}\bigtriangleup^{1}_{k}-M_{0}\xi_{k}^{+}) (\mid \bigtriangleup^{2}_{k}\mid-\bigtriangleup^{2}_{2,k})=0. 
\end{split}
\end{equation}

\noindent The second equation in Eqn.(5) is met only when $ \textbf{k}=(0,0)$ and 
$(\pi,\pi) $. These two points give the same following condition,

\begin{equation}
16 t^{2}+M_{0}^{2}=(4t'+\mu)^{2}. 
\end{equation}

\noindent When the above condition (Eqn.(6)) is satisfied, the energy gap closes. 
Thus, we find the topologically trivial and the non-trivial regions as shown in Fig.2. We will explore the Chern number associated with such phases in the next section.
 
\begin{figure}
\includegraphics[scale=0.75]{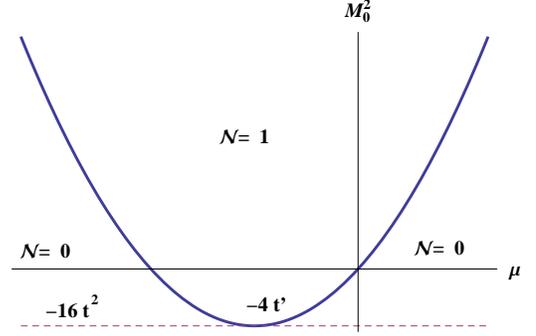} 
\caption{Schematic phase diagram of the coexistence phase of SC $ d_{x^2-y^2}+(p_x+i p_y) $ order and SDW order in the case where $k=(0,0)$ and $(\pi,\pi)$. 
The $x$-axis labels the parameter ${\mu}$  and the $y$-axis labels the parameter 
${M_{0}^2} $. Integer ${\cal N}$ labels the Chern number of the coexistence of SC and SDW.}
\label{fig.2}
\end{figure} 
 
\section{Phase diagram and the Chern number} 

To examine the topological phase transition(TPT), it is convenient to use the dual Hamiltonian instead of the original one. This can be done through a constant unitary 
transformation matrix $D$ as, 

\begin{eqnarray}
{\cal H}^{\rm D}({k})=D{\cal H}({k})D^{\dagger},
\quad
D=\frac{1}{\sqrt{2}}
\left(
\begin{array}{cc}
1 & i\sigma_y \\
-i\sigma_y & -1
\end{array}
\right),  
\end{eqnarray}

\noindent resulting 

\begin{equation}
\label{eq.2}
{\cal H}^{\rm D}({k})= \left( \begin{array}{cccc}
\xi_{k}^{+}+M_{0} & \bigtriangleup^{2}_{k}-\bigtriangleup^{1}_{k}  & 0 & \xi_{k}^{-} \\
\bigtriangleup^{2\ast}_{k}-\bigtriangleup^{1}_{k} & -\xi_{k}^{+}-M_{0} & -\xi_{k}^{-}  & 0 \\
0 & -\xi_{k}^{-} & -\xi_{k}^{+} +M_{0} & -\bigtriangleup^{2\ast}_{k}-\bigtriangleup^{1}_{k}\\
\xi_{k}^{-} & 0 & -\bigtriangleup^{2}_{k}-\bigtriangleup^{1}_{k} & \xi_{k}^{+}-M_{0}
\end{array} \right).
\end{equation} 

\noindent In the limit $t\rightarrow 0$, the dual Hamiltonian ${\cal H}^{\rm D}({k})$ 
in the leading order around ${\bf k}=(0,0)$ and $(\pi,\pi)$ gives rise to the following two $2 \times 2  $ block-Hamiltonian as, 

\begin{widetext}
\begin{equation}
\label{eq.2}
{\cal H}^{\rm D}({k})= \left( \begin{array}{cccc}
 -4 t'-\mu +M_{0} &  \bigtriangleup_{0}^{2}(k_{x}+i k_{y})  & 0 &  0 \\
\bigtriangleup_{0}^{2}(k_{x}-i k_{y}) &  4 t'+\mu -M_{0} &  0  & 0 \\
0 &  0 & 4 t'+\mu +M_{0} & - \bigtriangleup_{0}^{2}(k_{x}-i k_{y})\\
0 & 0 & - \bigtriangleup_{0}^{2}(k_{x}+i k_{y}) & -4 t'-\mu -M_{0}
\end{array} \right).
\end{equation} 
\end{widetext}

We notice here that the above dual Hamiltonian have a close similarity to the Hamiltonian 
of the spinless chiral $ p+ ip $ supercondutor discussed in\cite{read00}.

In order to study the phase diagram of this Hamiltonian one needs to determine the phase boundaries corresponding to gapless regions since the topological invariants can not change without closing the bulk gap. For the present model, the critical lines are 
determined by solving equation(8), i.e., $ M_0=\pm (4t'+\mu)$ for the 
upper(lower) blocks in the case with $\textbf{k} =(0,0)$ and $(\pi,\pi)$. The phase becomes  topological in the region when $ M_0<\pm (4t'+\mu)$ whereas it is trivial for $ M_0>\pm (4t'+\mu)$. 

Similarly, one can draw a phase diagram for any $ \textbf{k} $ point in the RBZ. e.g. for
 $\textbf{k} =(0,\pi)$ and $(\pi,0)$, we get the same condition for the topological phase transition to occur from second condition of Eqn.(5), i.e. $(4t'-\mu)\neq 0, \bigtriangleup^{1}_{0}=0$. This leads to the following 
 Eqn. of the critical lines $ M_{0}^{2}=(4t'-\mu)^{2} $. The topological phase exist in the region $ M_0<\pm (4t'-\mu)$ whereas it is trivial for $ M_0>\pm (4t'-\mu)$. 
 
Based on the finiteness of the Chern number given below, we propose a phase diagram in the 
in the former case (Fig. 2) in the $(\mu, M_0^2)$ plane which distinguishes the topological and non-topological phases. This is the new result of the present manuscript. 
Thus, it is obvious that both these phases are separated by a quantum phase transition 
line. The Chern numbers in the topological phases of Fig. 2 are calculated below.   
 

It is well known that the topological phases can be characterized by  ${\it Chern}$ 
${\it numbers}$. For a specific model Hamiltonian $h(k)=\sum_{\alpha}d_{\alpha}(k)\sigma_{\alpha}$, with $\sigma_{\alpha}$, the Pauli matrices and $d_{\alpha}(k)=[d_1(k), d_2(k), d_3(k)]$, the Chern number can be calculated from the expression

\begin{equation} 
\label{eq.5}
{\cal N}=\frac{1}{4\pi}\int d^2k\: {\hat d(k)}\cdot(\frac{\partial{\hat d(k)}}
{\partial k_x}\times\frac{\partial {\hat d(k)}}{\partial k_y}), 
\end{equation}

\noindent where the unit vector $\hat{d}(k)={\textbf{d}(k)}/\sqrt{\sum d^2(k)}$ 
characterizes a map from the Brillouin zone vector $k$ to unit sphere. The Chern number 
simply counts the number of times $\hat{d}(k)$ wraps around the unit sphere as a function 
of $k$. In the present model the Chern number for the case $ M_0<\pm (4t'+\mu)$ is 
calculated as ${\cal {N}}=1$ whereas it vanishes when $ M_0>\pm (4t'+\mu)$.  
In presence of SDW order, the $p+ip$ SC state has odd parity and $s_z$=0 symmetry 
which is a fully gapped system. Due to SDW order, it has $U(1)$ spin rotation along say, 
$z$-axis and $\pi_0(C_2)=Z$ which corresponds to class $A$ in accordance with the symmetry classification of Altland and Zirnbauer\cite{altland97}. This means that there are infinite number of distinct topological SC with $s_z$ conservation and are labeled by an integer 
which is the Chern number\cite{lu14}. This is associated with the number of chiral fermion edge modes. 
 
\section{Edge states and the Vortex structure}
In order to see the evolution of edge states in this model(coexistence of $p+ip$ SC 
and SDW order),   we studied it numerically on a cylindrical geometry with periodic 
boundary condition in $y$-direction and open boundary condition in $x$-direction.  
We solved the eigen-value problem where the Hamiltonian has 
been diagonalized on $N_x$=100 sites. The energy dispersion $E_k$ versus $k_y$ has been obtained and hence the edge states has been shown in Fig. 3. As it is already been 
discussed in the previous section, two chiral edge states characterize the topological 
phase in this model. 

\begin{figure}
\includegraphics[scale=1]{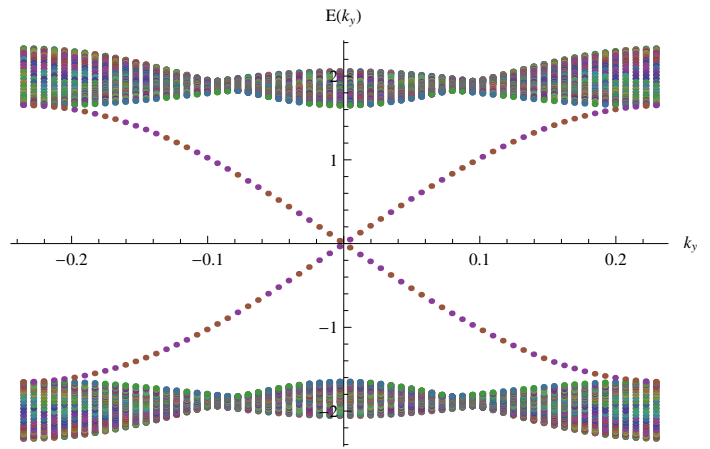} 
\caption{Edge-state spectrum of the coexistence phase of $ p_x+i p_y $ SC
in presence of SDW order on a cylindrical geometry. Parameters are chosen as, 
$ t'=-0.30t $ eV, $\bigtriangleup_{0}^{2}=t $, $ \mu=.25t $, $ M_{0}=.3t $ for a 
lattice of $ N_{x}=100 $ sites.}
\label{Fig.3}
\label{Fig.3}
\end{figure} 

It is well known that the vortex of a topological SC with odd topological quantum 
number ${\cal N}$ carries an odd number of Majorana zero modes. The existence of such 
zero modes in the vortex core of a $p+ip$ SC is shown to be due to index theorem\cite{tewari07,sato09}. In the present case, since the SC is coexisting with 
SDW order, the topological classification is always trivial for class $A$ in 1D. Since  
the existence of the zero modes in the vortex core is determined by the symmetry classification in one space dimension less, the topological $p+ip$ SC with the chiral 
edge modes won't support such zero energy vortex bound state\cite{lu14}. 
 
\section{Conclusion}
In conclusion, we summarize the main findings of the present manuscript. 
We consider a possible   
coexistence of singlet SC and SDW which induces a triplet SC component as well. 
The singlet SC is taken to be of $ d_{x^2-y^2}$  symmetry whereas the SDW order 
parameter is of s-wave and the triplet SC is $p+ip$ type symmetry. Such a Hamiltonian 
is shown to yield a non-trivial coexistence phase which is topological in addition to 
the conventional one. A phase diagram characterizing different topological phases is 
construted. The Chern numbers and hence the nature of the topological phases are 
determined. The edge state spectrum and the possibility of whether the vortex state harbouring the zero modes are discussed.
\label{}
 
\section{Acknowledgements}
This work is supported by Council of Scientific and Industrial Research (CSIR), India.  
  




\end{document}